\documentclass[fleqn,usenatbib]{mnras}

\usepackage{newtxtext,newtxmath}
\usepackage[T1]{fontenc}
\usepackage{ae,aecompl}
\usepackage{graphicx}	
\usepackage{amsmath}	
\usepackage{amssymb}	
\usepackage{footnote}

\title{Optimization of Radio Array Telescopes to Search for Fast Radio Bursts}

\author[Peterson J. B. et. al.]{
Jeffrey B Peterson$^{1}$\thanks{E-mail: jbp@cmu.edu },
Kevin Bandura$^{2,3}$\thanks{E-mail: kevin.bandura@mail.wvu.edu },
Pranav Sanghavi$^{2,3}$ \thanks{E-mail: pranav.sanghavi@mail.wvu.edu}
\
\\
$^{1}$McWilliams Center for Cosmology, Department of Physics, Carnegie Mellon University, 5000 Forbes Ave, Pittsburgh PA 15213 USA\\
$^{2}$LCSEE, West Virginia University, Morgantown, WV 26505, USA\\
$^{3}$Center for Gravitational Waves and Cosmology, West Virginia University, Morgantown, WV 26505, USA
}

\date{Accepted XXX. Received YYY; in original form ZZZ}

\pubyear{20}

\begin{document}
\label{firstpage}
\pagerange{\pageref{firstpage}--\pageref{lastpage}}
\maketitle

\begin{abstract}
We present projected Fast Radio Burst detection rates from surveys carried out using a set of hypothetical close-packed array telescopes. The cost efficiency of such a survey falls at least as fast as the inverse square of the survey frequency. There is an optimum array element effective area in the range 0 to 25 $\rm{m^2}$.  If the power law index of the FRB integrated source count versus fluence $\alpha = d ~ln R/d ~ln F > -1$ the most cost effective telescope layout uses individual dipole elements, which provides an all-sky field of view. If $\alpha <-1$ dish arrays are more cost effective.

\end{abstract}


\section{Introduction}
Fast Radio Bursts are bright millisecond radio flashes broadly distributed across the sky. Most FRB pulses arrive with dispersion measure much larger than the estimated Galactic contribution along the line of sight of detection, therefore FRBs are believed to originate in extra-galactic sources.
Four FRBs \citep{2017ApJ...834L...7T,frbcat} have been localized and are coincident with galaxies at redshifts of 0.19, 0.32, 0.48 and 0.66, reinforcing the extragalactic origin hypothesis.
Efforts to explain the exceptionally high inferred isotropic brightness temperature of these sources ($10^{38} \rm{~K}$) along with the all-sky rate ($\sim$2000/year above fluence 1 Jy ms) have employed extreme astrophysical models \citep{review1, review2}. If the few-millisecond duration of the bursts reflects the spatial extent of the source region, these cosmologically distant sources have an angular diameter smaller than $10^{-21}$ radians. The small source size allows FRB spectra to exhibit non-dispersive two-path interference when microlensed \citep{2019BAAS...51c.420R}, as well as scattering and scintillation that does exhibit dispersion as the pulses transit the intervening plasma. FRBs can therefore serve as a unique new probe of the microlensed and ionized universe.

The optimum parameters for a custom FRB telescope may depart substantially from previous telescope designs.  Below we examine a set of hypothetical telescope layouts and calculate the dependence of FRB detection rate on array element collecting area as well as survey frequency and number of elements. 

We restrict our analysis to close-packed arrays. Alternatives not considered include: Large single dishes with a focal plane array of feeds--these are more expensive than close packed arrays of the same collecting area; Dilute dish arrays--these have increased processing costs, with a different signal processing system than discussed here.  
 
\section{Projected FRB detection rates}

We consider two types of array elements: on-axis paraboloidal dishes and square aperture array tiles.  Each tile consists of $m=1, 4, 9 ...$ dipole-like antennas, analog summed to form a single element with one beam using $m$ antennas. For either type, $n$ elements are close-packed to form the telescope. We leave cylinder arrays \citep{chimesystem} to future analysis since these are hybrids the other types and should produce rates that fall between the two we consider.

Each element covers the same instantaneous field of view $\Omega=4\pi/G$, defined in terms of the peak antenna gain $G$ (\cite{era}. The effective collecting area of each element $A_{\rm{el}}$ is related to $\Omega$ via the diffraction relation, 

\begin{equation}
		A_{\rm{el}} ~\Omega =\lambda^2.
\end{equation}

Diffraction forces a trade-off: while an increase of $A_{\rm{el}}$ allows dimmer FRBs to be detected, the resulting increase in detection rate is moderated by a decreased field of view $\Omega$. Note that the total effective area of the telescope is $ A_{\rm{tel}}= 2 n A_{\rm{el}}$. The factor of two accounts for the assumption that both polarizations are used.

The beam solid angle $\Omega$ is limited by the horizon to half the sky. Furthermore, no antenna design allows uniform sensitivity across a full hemisphere, so a practical maximum to the field of view is  $\Omega \sim \pi$. The aperture array tile with $m=1$ is considered the `all-sky' option since it has the widest beam and also maintains substantial sensitivity from horizon to horizon.

The individual antennas within an aperture array tile are assumed to be dipole-like antennas close to a ground plane. These might be four-squares, clovers, sloping bow-ties, etc.  We generically assume $G=4$ and $\Omega = \pi$ for these individual antennas.

We assume the FRB detection rate can be represented as a power law 

\begin{equation}
	R(>F_{min}) = R_o \left[ \frac{F_{min} }{F_o} \right] ^\alpha  ~\frac{\Omega}{4 \pi}, 
\end{equation}  

where the all-sky rate normalization $R_o$ is drawn from results of a previous survey with fluence threshold $F_o $. We discuss the limited knowledge of the all-sky rate below. The limiting fluence $F_{min}$ is defined below.
The power law index of the integrated source count distribution function $\alpha $ has the euclidean value $-3/2$ if the median FRBs  redshift is well below one, the co-moving FRB rate does not evolve substantially over the observed redshift range and the local universe has negligible density structure. This index can be difficult to measure accurately because it can be difficult to estimate survey completeness near the fluence limit. Worse, estimates of $\alpha$ are often made by comparing surveys, which may have differing RFI cuts and search algorithm efficiencies.  To accommodate a range of possible indices we consider $\alpha = -1.0, -1.5, ~-2.0$. 

The limiting fluence $F_{min}$ of a survey is given by the radiometer equation 

\begin{equation}
	F_{min} =  \frac{\rm{SNR}_{\rm{min}} ~k ~T_{sys} ~\tau_b}{2 n A_{\rm{el}} \sqrt{\Delta \nu ~\tau_{\rm{b}}}} .
\end{equation}
                
$\rm{SNR}_{\rm{min}}$  is the signal to noise ratio threshold, which is set by the survey designers,
${\rm k }$ is the Boltzmann constant, ${\rm T_{sys}} $  is the system temperature, $\tau_{\rm{b}}$ is the burst duration and $\Delta \nu =f  \nu_s $ is the bandwidth of the observation, which can be expressed as a fraction $f$ of the center frequency of the survey $\nu_s $.

Equations 1 to 3  combine to express scaling laws for the FRB detection rate,

\begin{equation}
R(>F_{min}) = \frac {R_o {F_o}^{-\alpha} c^2}{4 \pi} \left[  \frac{  \rm{SNR}_{\rm{min}} ~ kT_{sys} ~\tau_b}{ 2~\sqrt{f ~ \tau_{\rm{b }} } } \right]  ^{\alpha}A_{\rm{el}} ^{-\alpha-1} ~{\nu_s}^{-\alpha/2-2} ~{n^{-\alpha}}.
\end{equation}

The quantities in square braces are consider fixed in this analysis.  
For the frequency range we consider ${\rm T_{sys}}$ is determined by receiver noise
for much of the sky, but at frequencies below
about 400 MHz a term should be added to ${\rm T_{sys}}$ to account for Galactic synchrotron emission.
Absent cost considerations, for $\alpha \sim -1.5$, equation 4 pushes the design in the direction of increased element area, lower frequency and higher element count. When costs are considered, below, the optimization is more complicated.

\begin{figure}
   \includegraphics[width=1\linewidth,height=0.82\linewidth]{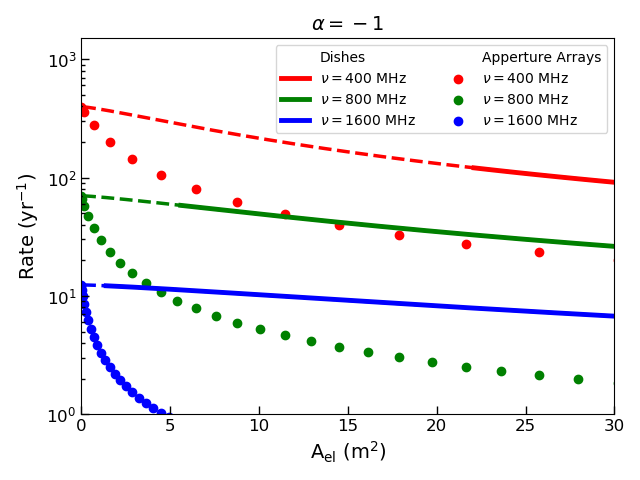}

   \includegraphics[width=1\linewidth,height=0.82\linewidth]{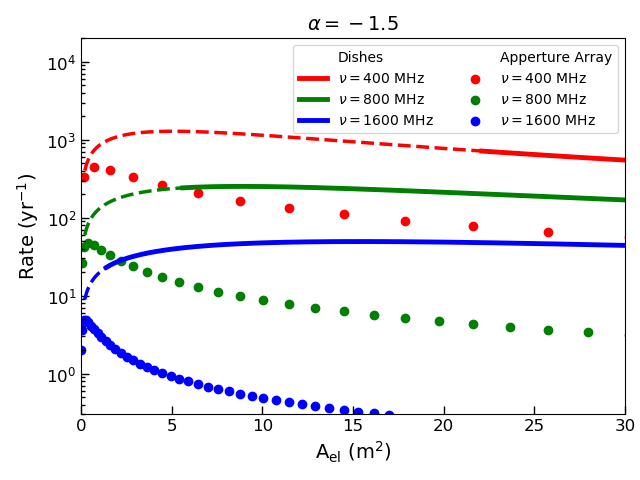}

   \includegraphics[width=1\linewidth,height=0.82\linewidth]{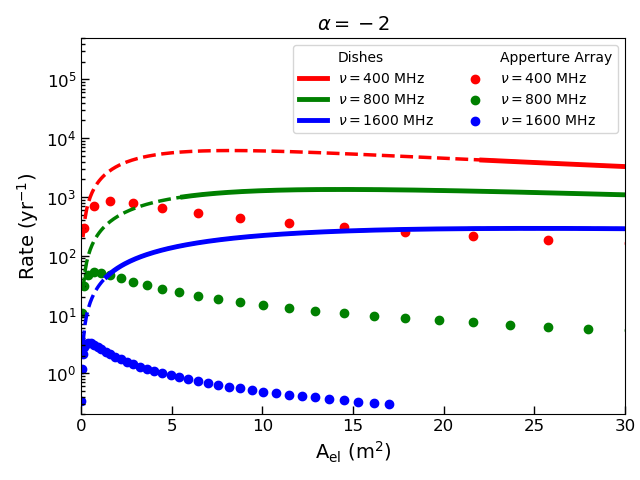}

\caption[Two numerical solutions]{Projected FRB detection rate, for $\alpha=-1.0, -1.5, -2.0$ with three survey center frequencies: 400 MHz (red), 800 MHz (green), 1600 MHz (blue) and two element choices: Dishes (smooth curves) and Aperture Array Tiles (solid circles).   Integer steps of aperture array dimension 1x1, 2x2, 3x3, etc. are shown.
   On-axis dishes smaller than $ d= 5 \lambda$ are not commonly used because of feed blockage inefficiency, so these rates are shown as dashed curves. Olga Navros helped prepare this figure.}
\end{figure}

\section{Cost Model}

We divide construction costs into two components: 1) The Radiation Collector--these cost increase with $A_{\rm{el}}$; 2) The Signal Processor--these cost are independent of $A_{\rm{el}}$. 

For dish arrays we adopt Radiation Collector cost function $C_d = D_o A_{\rm{el}}^{1.25} ~n $, a bit steeper than a linear function of dish area \citep{10.1117/12.552181}.
We assume the dish cost is independent of frequency, since many commercially available dishes are more precise than we require at these low frequencies.

For aperture array tiles we adopt Radiation Collector cost function $C_a = A_o ~m ~n$, where $A_o$ is the cost of an individual dipole-like antenna (\cite{clover}).

For the Signal Processor we adopt the cost function $C_s = S_o ~f ~\nu ~n$.  This assumes all signal processing operations required, such as digital beam forming and de-dispersion, can be accomplished using algorithms with computational asymptotic order $N$ or $N ~ln N$.  We absorb slowly varying $ln N$ factors into the normalization $S_o$.

To set the cost coefficients $D_o$, $A_o$ and $S_o$ we use cost data for an FRB-search array of 12 six-meter dishes we built at Green Bank WV in 2019.   The ratio of the three cost components is Dish : DSP : Dipole  = 1 : 0.92 : 0.067,  with $\nu_s=600\rm{~MHz}$, $f= 0.66$, which translates to relative cost coefficients 
$D_o= 0.029 [\rm{m^{-2.5}}]$, $ S_o = 2.3 \times 10^{-3}\rm{~[MHz^{-1}}]$ and $A_o = 0.067$.

\section{Frequency Dependence}

Projected FRB detection rates fall dramatically with frequency, as seen in Figure 1, primarily because of the $\lambda^2$ factor in equation 1.
Note that our calculations show only the instrumental impact of telescope parameters on the rate, while the on-sky rate also depends on the (unknown) source spectrum and on possible pulse broadening during propagation. This scatter-broadening is expected at low frequencies due to plasma structure in the host galaxy or in the Milky Way and may produce a low frequency falloff of the rate. The observational data on scattering are ambiguous. About 5 \% of the entries in FRBCAT \citep{frbcat}  have a measured scattering spectral index ($d ~ln ~\tau_{\rm{scat}}/ d ~ln ~\nu$) with values ranging from --3.6 to --4.8, consistent with scattering in turbulent plasma. However there are also FRB spectra that show no measurable scatter broadening, even at 400~MHz, the lowest detection frequency so far \citep{chime1}. 

We summarize current published rate estimates in Table 1 and Figure 2. FRBs have been detected from 400~MHz to 8~GHz. Most archival searching has been done near 1.5~GHz so there is substantial rate information at that frequency. The CHIME team reports numerous FRB detections at their lowest frequency 400~MHz but the team has not published an all-sky rate \citep{chimesystem, chime1, chime2}. At and below 327~MHz there are reports of non-detection but the fluence limits of these surveys are much higher than for the higher frequency surveys. So far, there is at best weak evidence for any spectral slope of the FRB detection rate.

Since we calculate the instrument-specific impact of telescope parameters on the detection rate, we are implicitly assuming a flat prior distribution for $R_o(\nu_s)$, which is consistent with current data.  

\begin{figure}
 \includegraphics[width=\columnwidth]{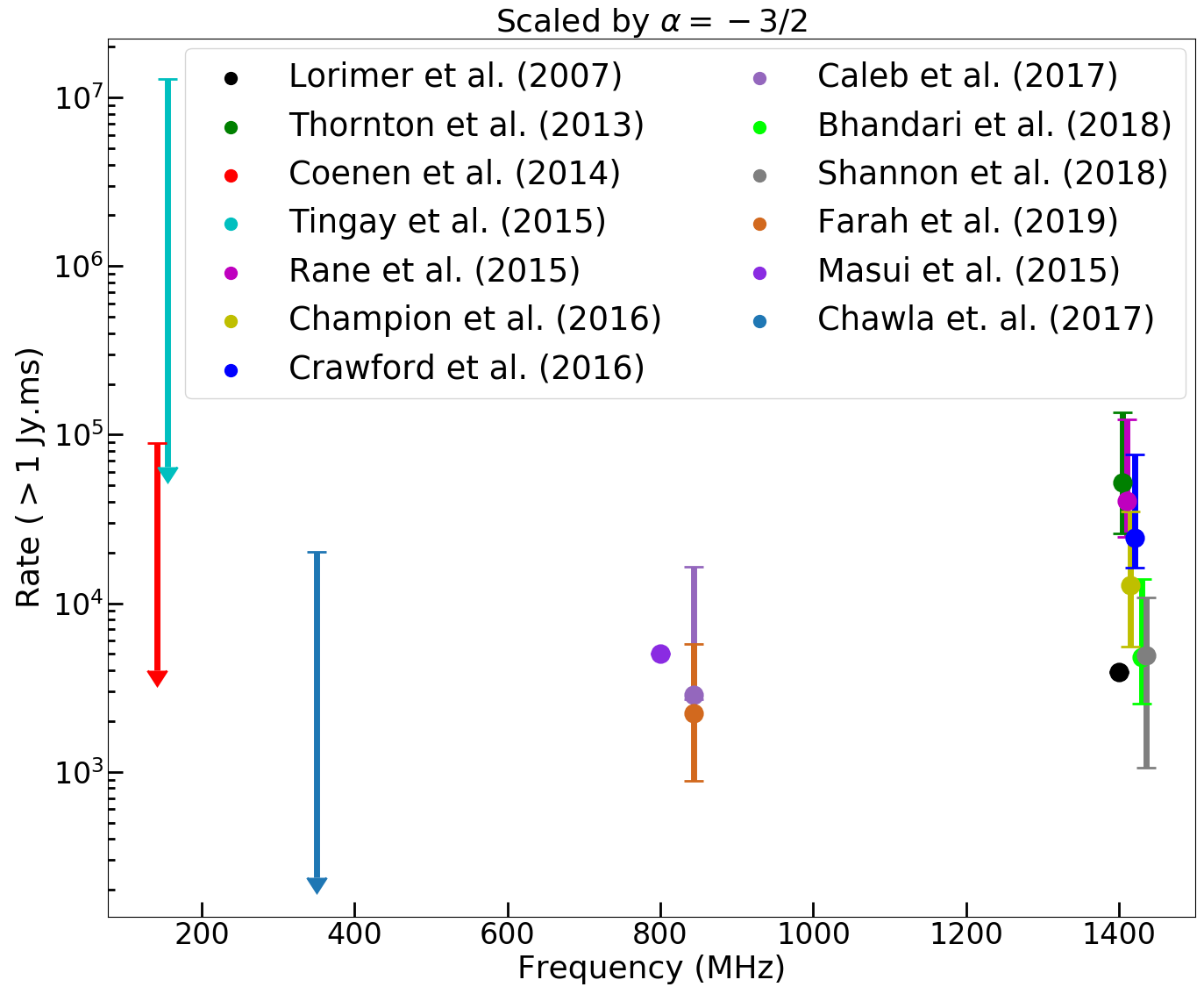}
 \caption{ Extrapolated all-sky integral FRB rates. Rate data from Table 1 have been adjusted to a common fluence threshold using  $\alpha$ = -3/2. The results are offset in frequency by 5 MHz each to show evolution of the claimed rate with time. There is no clear trend with frequency.}
\end{figure}

\begin{table*}[t]
\caption{Published estimates of the all-sky FRB rate above the survey fluence threshold ($\rm{F_{lim}}$). Survey center frequencies are shown.}
\label{published rates}
{\renewcommand{\arraystretch}{1.5}
\begin{tabular}{lccc}
\hline
Survey                              & Freq[MHz]  & Rate [events/sky/day]       & $F_{\rm{lim}}$ [Jy.ms]                                                  \\ \hline
 Lorimer Burst \citep{Lorimer}  & 1400             & 225                       & 6.7                       \\
 Parkes 2013 \citep{parkes2013}     & 1400             & $10000^{6000}_{-5000}$    & 3                         \\
 LOFAR PFRT 2014 \citep{lofar2014}& 142              & $< 150$                    & 70    \\
 MWA 2015  \citep{mwa2015}      & 155              & $< 700$                      &  700                  	\\
 GBT 2015  \citep{gbt2015}      & 800              & $5000$                    & 1                         \\
 Parkes 2016 \citep{parkes20160} & 1400             & $4400^{+520}_{-310}$       & 4                         \\
 Parkes 2016 \citep{parkes20161}     & 1400             & $7000^{+5000}_{-3000}$    & 1.5                       \\
 Parkes 2016   \citep{parkes20162}  & 1400             & $3300^{+3700}_{-2200} $    & 3.8                       \\    
 GBNCC 2017 \citep{gbncc2017}   & 350               & $ < 3620$  & 3.15                                         \\
 UTMOST 2017  \citep{utmost2017}   & 843              & $5.0^{+18.7}_{-4.7}$     & 69                        \\
 Parkes 2018  \citep{parkes2018}  & 1400             & $1700^{+1500}_{-900}$      & 2                         \\ 
 ASCAP 2018    \citep{ascap2018}  & 1400             & $37 \pm 8 $               & 26                    	\\
 UTMOST 2019   \citep{utmost2019}  & 843              & $98^{59}_{-39}$           & 8                         \\    
 \hline
\end{tabular}}
\end{table*}

\section{Number of array elements}

Budgets have limits, so we fix the total telescope cost at the arbitrary value 2000 units, which allows for an interesting detection rate of 100s to 1000s per year.  Under this constraint increasing the element collecting area will entail a reduction of the number of elements. This creates a peak in the rate function $R(A_{\rm{el}})$ at which the cost efficiency is maximized.

\section{Element Effective Area}

The dependence of projected FRB detection rate on $A_{\rm{el}}$ is shown in Figure 1. The assumptions used in this plot are: $T_{\rm{sys}}= 50~\rm{K},~ \tau_b = 2~\rm{ms},~ SNR_{min}=10$ and $f=0.66$, We used the most recent Parkes Survey to normalize the rate: $R_o = 1700 \rm{/yr} , F_o= 2~\rm{Jy~ms}$. This is a conservative choice since other surveys find substantially higher rates. We posted the spreadsheet and python notebooks used to make these calculations and plots on GITHUB \footnote{\texttt{https://github.com/WVURAIL/Optimization-of-Radio-Array-Telescopes-to-Search-for-Fast-Radio-Bursts} } so others can calculate rates under different assumptions. 
 
\section{Discussion}

For aperture arrays the optimal rate decreases by 0.062 per frequency doubling for $\alpha =-2$ and 0.20 per frequency doubling for  $\alpha =-1$.  For dish arrays the decrease is 0.22 per frequency doubling ($\alpha =-2$) and 0.18 per frequency doubling ($\alpha =-1$). All these factors are less than 1/4 so we find the cost efficiency of such telescopes falls with frequency with at least two powers of $\nu_s$. Once the on-sky rate versus frequency is measured, if low frequency FRBs are rare, the source spectral slope and/or scattering spectral slope may compensate for the telescope cost efficiency spectral slope. The optimum frequency can then be determined by locating the point where the combined slope is zero. Published rate data are not sufficiently precise to determine this optimum frequency. However, scattering is absent in several published FRB spectra at 400~MHz, indicating that a high throughput search below this frequency should yield FRB detections. 
If the CHIME team finds an all sky rate versus frequency with spectral index greater than 2 they may be able to locate the optimum frequency within their band. If the index is below 2 the optimum lies below 400 MHz. 

The source count index $\alpha$ determines which array type (dishes versus aperture array tiles) is most cost efficient. The critical point is near $\alpha = -1.0$.  If dim FRBs are rare enough ($\alpha > -1$) the all-sky array is the the most cost efficient choice.

In the future it is reasonable to assume signal processing costs will continue to fall. As this happens the cost efficiency of the all-sky aperture array will see the greatest improvement to cost efficiency, since this configuration has costs strongly dominated by the signal processor.

Paying attention to cost efficiency allows the telescope designer to use funding efficiently, however there may be science goals that push the design away from peak rates shown in Figure 1. For example, widening the field of view can be useful to increase the event rate of rare, bright, nearby FRBs. 
Some emission models \citep{2019MNRAS.485.4091M} predict weak x-ray to gamma-ray afterglows which would only be detectable for nearby FRBs. The most constraining test of these models will come from an all-sky FRB telescope. 

Telescopes with element collecting area larger than the area at the rate peak may also have a scientific benefit that justifies increased cost. Increasing the collecting area means the survey will be deep rather than wide, which increases the average redshift of the FRBs detected, providing longer paths on which to study the ionized universe and search for microlensing, while also allowing study of cosmic evolution of the FRB event rate. 

Throughout this analysis we have focused on the FRB detection rate using a single close-packed array telescope, setting aside the important topic of precise localization of the sources. Outrigger arrays will be needed to provide this localization. These can be built of the same elements as the central close packed array, and placed hundreds to thousands of kilometers away.   The outriggers can have a combined area substantially smaller than the central array since the central array provides high SNR waveform templates which can be used to recover the weaker signal from the outriggers.

When designing a telescope specifically to detect FRBs some design constraints can be relaxed.  Array telescopes can have aliases, false point source locations. The position of these aliases move on the sky with frequency while the true source position remains fixed, so for FRBs which sweep in frequency, the aliased positions can be identified and deleted. This means the dipole-like antennas can have spacing wider than the Nyquist spacing needed to eliminate aliases. This increases $A_{\rm{el}}$. Compared to intensity mapping telescopes, FRB telescopes also have less stringent requirements for gain calibration precision and low side lobe response.

\section{Conclusion}

We have presented optimization calculations for radio telescope arrays used for the detection of Fast Radio Bursts.  The instrument-specific detection rates fall with frequency indicating low survey frequencies are preferred. However, other factors influence the on-sky rate, such as possible low frequency falloff of rates due to scattering, and the intrinsic source spectral index. These other factors are currently poorly constrained. Current published data is inadequate to identify an optimum survey frequency. The lack of scattering at 400~MHz in some FRB spectra indicate searches at frequencies below 400~MHz should be productive.

If the source count index $\alpha > -1$, cost efficiency favors the minimal element area--the all-sky aperture array designs.  For $\alpha < -1$, cost efficiency favors dish arrays over aperture arrays, with the peak in cost efficiency moving to larger element area for smaller $\alpha$. 

Apart from cost considerations, the science goals for wide and narrow field observation differ.
All-sky telescopes can be used to understand the FRB emission mechanism, by allowing detection of rare, nearby, bright events which can be followed up at other wavelengths with high sensitivity and spatial resolution. In contrast, larger element area allows detection of dimmer FRBs along longer paths through the ionized universe, allowing improved constraints to the evolution of FRB rates over cosmic time, increased occurrence of lensed events and tighter cosmological constraints.




\bibliographystyle{mnras}
\bibliography{papers.bib} 



\label{lastpage}
\end{document}